# Redesigning Commercial Floating-Gate Memory for Analog Computing Applications

F. Merrikh Bayat[1], X. Guo[1], H. A. Om'mani[2], N. Do[2], K. K. Likharev[3], and D. B. Strukov[1*]
[1] UC Santa Barbara, Santa Barbara, CA 93106-9560, U.S.A.
[2] Silicon Storage Technology Inc., A Subsidiary of Microchip Technology Inc., San Jose, CA 95134, U.S.A
[3] Stony Brook University, Stony Brook, NY 11794-3800, U.S.A.
*Corresponding author; email strukov@ece.ucsb.edu

***Abstract*— We have modified a commercial NOR flash memory array to enable high-precision tuning of individual floating-gate cells for analog computing applications. The modified array area per cell in a 180 nm process is about 1.5 µm$^2$. While this area is approximately twice the original cell size, it is still at least an order of magnitude smaller than in the state-of-the-art analog circuit implementations. The new memory cell arrays have been successfully tested, in particular confirming that each cell may be automatically tuned, with ~1% precision, to any desired subthreshold readout current value within an almost three-orders-of-magnitude dynamic range, even using an unoptimized tuning algorithm. Preliminary results for a four-quadrant vector-by-matrix multiplier, implemented with the modified memory array gate-coupled with additional peripheral floating-gate transistors, show highly linear transfer characteristics over a broad range of input currents.**

## I. Introduction

Nonvolatile floating gate memory devices are very attractive for analog computing [1, 2], because their state may be tuned continuously. For example, vector-by-matrix multiplication is a bottleneck in many signal processing and artificial neural network tasks [1, 3]. Floating-gate devices enable such multiplication, for slowly changing matrix elements, with relatively low precision, but very high performance: high speed, high density, and low power [4-6].

The so-called synaptic transistors and similar devices [4, 5, 8-17, 19] (see also reviews [18, 7]) is the most commonly reported implementation of this idea. This technique is very convenient due to its compatibility with the generic CMOS fabrication process. Its handicap is a large cell area and lower quality (e.g., in terms of retention [20]) of the floating-gate devices in comparison with those used in highly optimized flash memories. Figure 1 illustrates one such flash memory technology (ESF-1), from Silicon Storage Technology, Inc. (SST) [21]. Its relative cell area $A/F^2$, where $F$ is the half-pitch of the employed CMOS process, is close to 20. Such relative area is at least an order of magnitude smaller than that in synaptic transistors [8-10, 15, 17]. However, the baseline SST floating-gate technology has been designed for digital NOR flash memory applications, and does not allow setting a precise analog state of each cell, necessary for analog applications. This paper describes a successful redesign of the SST memory, which enables such individual cell tuning.

This work was supported by DARPA under Contract No. HR0011-13-C-0051UPSIDE via BAE Systems. Useful discussions with M. Graziano, O. Kavehei, L. Sengupta and V. Tiwari are gratefully acknowledged.

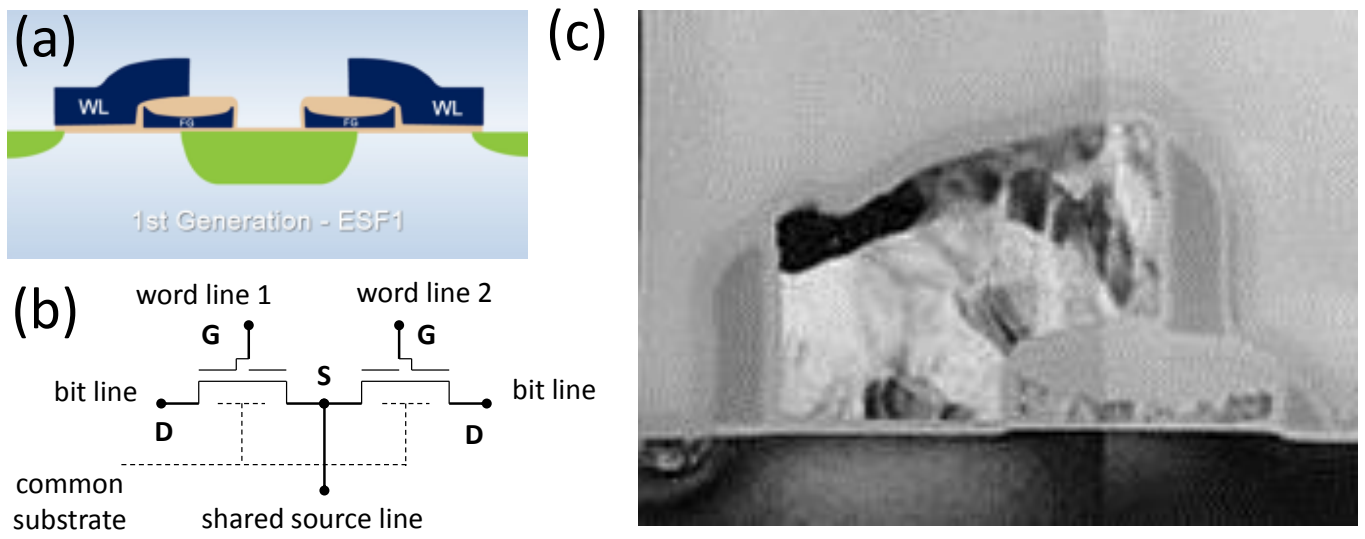

Fig. 1. SST's ESF-1 technology [21]: (a) schematic cross-section of a supercell, (b) its equivalent circuit, and (c) TEM cross-section image of one half of the supercell implemented in a 180-nm process.

## II. Memory Array Design

The SST NOR memory array consists of "supercells" (Fig. 1). Each supercell is a common-source assembly of two floating-gate memory cells with a highly asymmetric structure: the control gate (usually connected to a "word" line) overlaps the drain region of cell's MOSFET transistor, while being separated from its source region by the floating gate. Because of that, the direct effect of the gate voltage on the process of electron emission by the source is very small. This is evident from the readout characteristics of the cell, shown in Fig. 2: at $V_{DS} > 0$, when the source-to-drain current is due to the electron emission from the source, a large gate voltage is necessary to open the transistor of a fully programmed cell (with negatively charged floating gate). On the other hand, at $V_{DS} < 0$, when electrons are emitted by transistor's drain, the effect of control gate voltage on the current is much stronger, while that of the floating gate charge is much weaker.

The same structure asymmetry affects the switching dynamics of the cell (Fig. 3). During the "programming" process, the negative charge of the floating gate may be increased very fast using very effective hot-electron injection from the source area of transistor's channel, while the simplest way to decrease it (and hence "erase" the cell) is via the Fowler-Nordheim tunneling of electrons from the floating gate to the control gate, by applying a rather high voltage ( ~ 11 V) to the latter electrode.

The top row of Fig. 4 shows the usual structure of the NOR flash memory and its programming/erasure voltage protocols, employing these properties of the SST cells. In this architecture, cells of the same row share transistor source and control gate ("word") lines, while transistor drains of all cells of the same column are connected to the same "bit" line. Fig. 4a shows the set of applied voltages used for programming of the top left cell, while avoiding state disturb in all other cells. In particular, a positive bias $V_D^{P'} > 2$V, applied to all unselected bit lines, inhibits unintentional hot-electron injection in all unselected cells, including type-A half-selected cells (sitting on the selected word line). Also, grounding of unselected word lines guarantees the absence of disturb processes (such as the back Fowler-Nordheim tunneling) in all unselected cells including half-selected cells of type B (sharing the source voltage with the selected cell). As Fig. 3a indicates, the same programming protocol, only with pulsed source voltage and slightly modified voltage values, allows analog programming of the selected cell, also without disturbing the half-selected cells, regardless of their charge state.

Unfortunately, in this memory architecture the opposite process of cell erasure (Fig. 4b) is much less controllable. Namely, the fully selected cell and the type-C half-selected cell share their gate and source voltages, and due to the cell structure (Fig. 1) the process responsible for erasure (the Fowler-Nordheim tunneling of electrons from the floating gate to the control gate) is only weakly affected by the drain voltage $V_D$ – the only voltage which may be different for these two cells. (The possible increase of $V_D^{E'}$ is limited by the onset of large drain-to-source current.) For digital applications this feature is not a handicap, because in flash memories all cells of the same row are erased simultaneously. However, in analog applications it is highly desirable to perform not only a gradual programming of each cell, but also a gradual erasure of each cell without disturbing its neighbors. Our detailed measurements (see, e.g., Fig. 3c,d) have shown that in the baseline architecture (Fig. 4a-c) the latter operation is impossible for any bias voltage set.

To resolve this problem, we have modified the array structure (without changing the optimized cell fabrication technology) as shown in the bottom row of Fig. 4, i.e. by re-routing the gate lines in the "vertical" direction, i.e. perpendicular to the source lines. A straightforward analysis of the data shown in Figs. 2 and 3 shows that the new design resolves the half-selected cell disturb problem, by using the applied voltage protocol shown in Fig. 4d,e, with $V_G^E \approx 8.5$V, $V_S^{E'} \approx 3$V, and $V_D^E \approx 3$V.

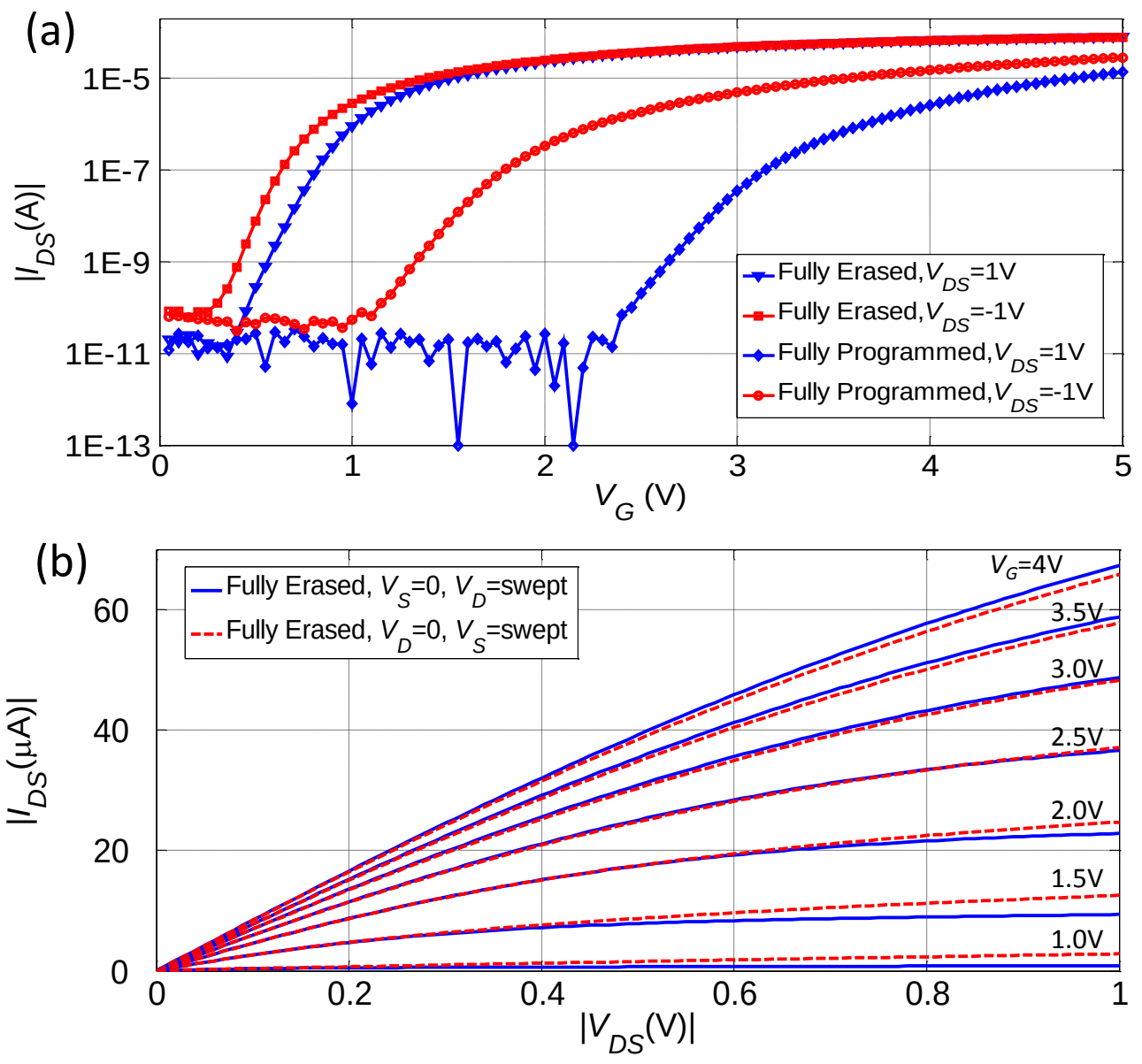

Fig. 2. Readout characteristics of 180-nm ESF-1 memory cells: Drain-source current as a function of (a) gate and (b) drain-sorce voltage.

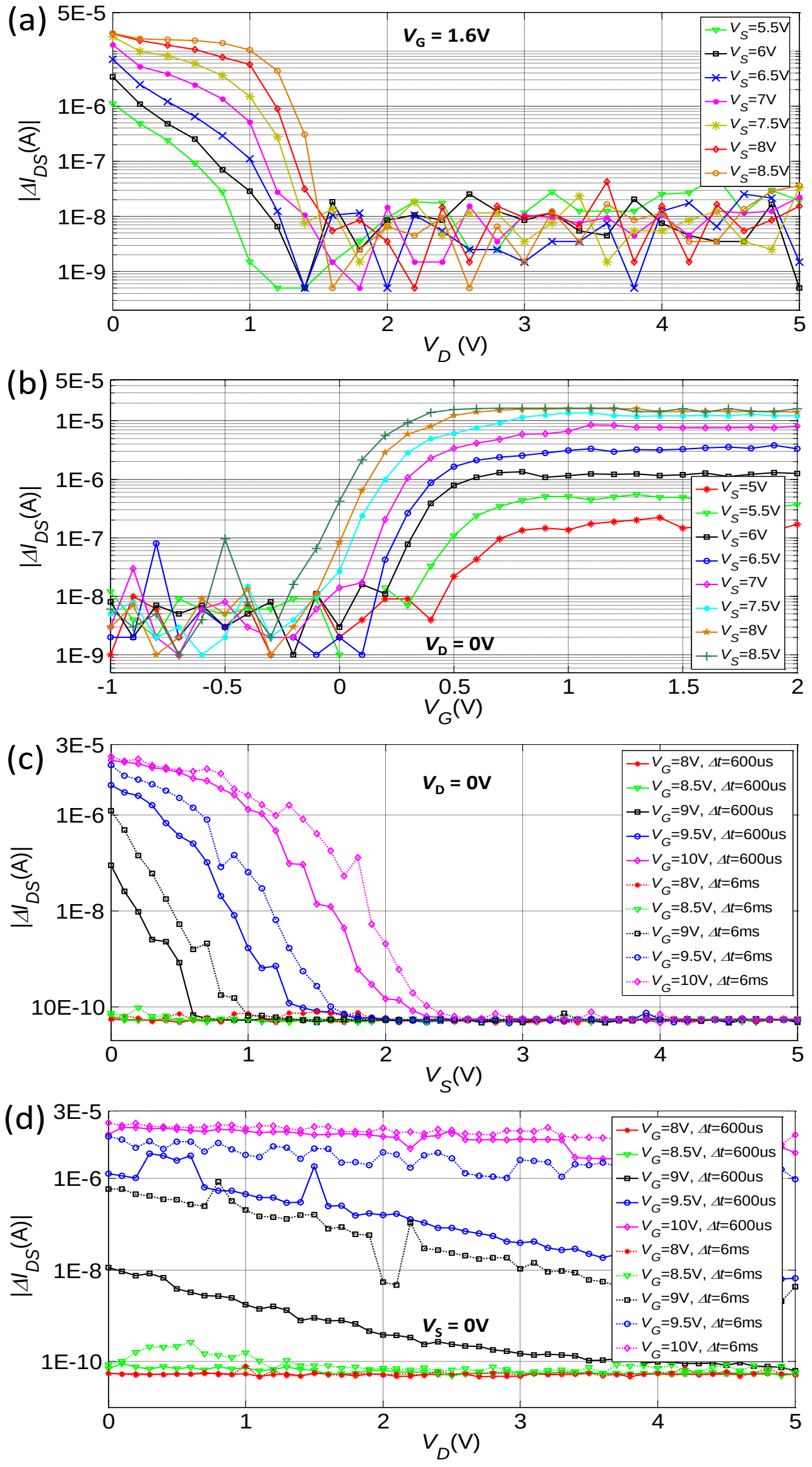

Fig. 3. Analog tuning of 180-nm ESF-1 memory cells, characterized by the change in source-to-drain current $I_{DS}$ (as measured at $V_G$ = 2.5V, $V_D$ = 1V, and $V_S$ = 0V) under effect of applied voltage pulses: (a, b) gradual programming of an (initially erased) device with 5-μs source voltage pulses of various amplitudes $V_S$; (c, d) gradual erasure of an (initially programmed) device with gate voltage pulses of various amplitudes $V_G$ and durations $\Delta t$.

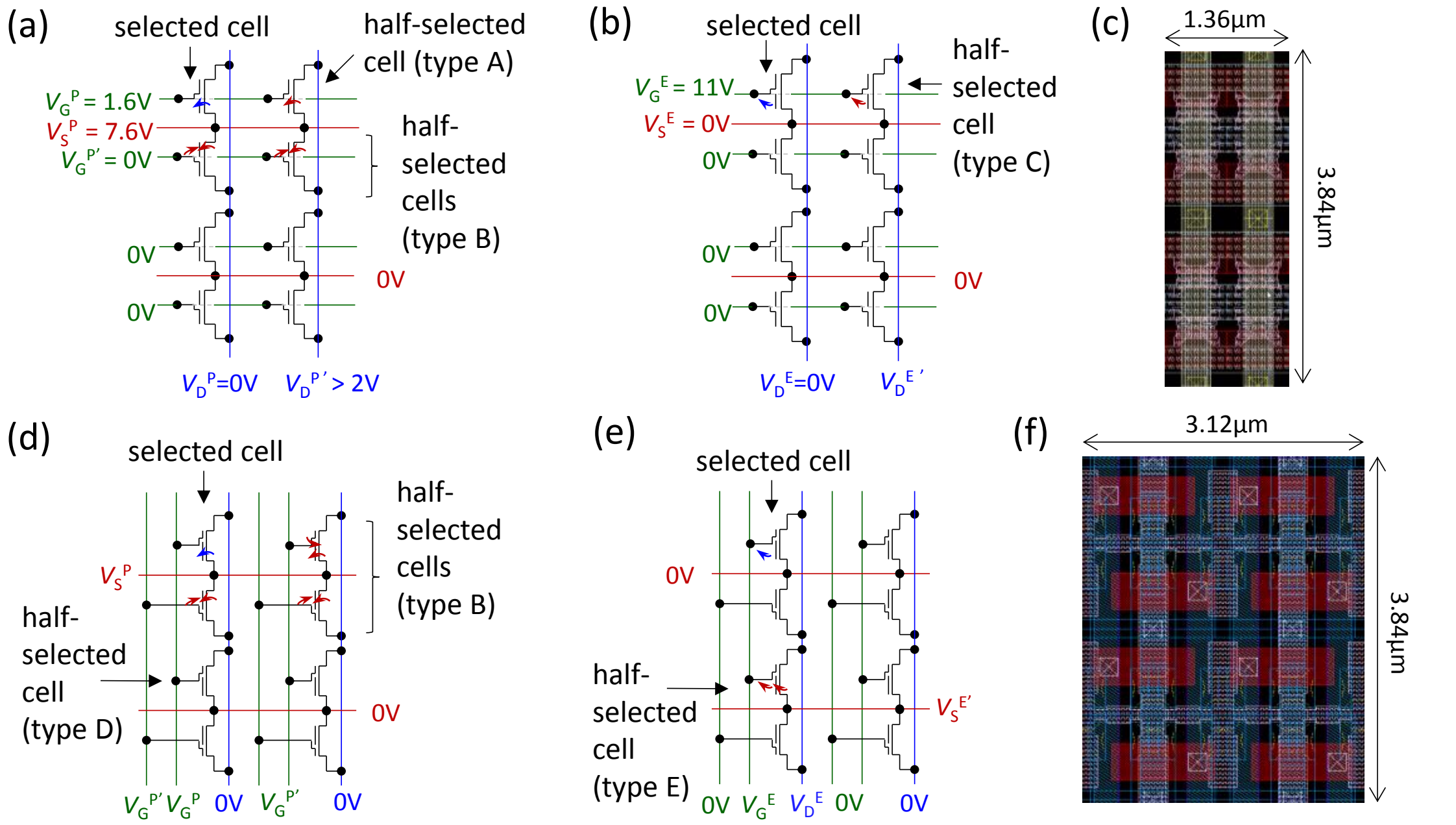

Fig. 4. Floating gate recharging effects: (a, b) – in the original SST array (c), and (d, f) – in the array with modified routing (f), on the example of a 2×2 supercell array fragment. Voltages shown on panels (a, d) correspond to programming of the top left cell, while those on panels (b, e), to its erasure. Blue and red arrows show, respectively, the useful and undesirable recharging processes. Line colors are for clarity only.

Indeed, for the programming operation, most of the half-selected cells are of the type B, while the disturb in type D cells, with $V_G^{P'}$ = -1V, is even less problematic. For the erase operation, the new gate line routing enables taking advantage of the very strong nonlinearity of possible Fowler-Nordheim and hot-electron tunneling currents (as functions of, respectively, the drain and source voltages), to completely inhibit these effects in all unselected cells including half-selected cells of type E.

## III. Experimental Results

The SST cell array with the architecture shown in Fig. 4d,e has been designed, fabricated (so far in the 180-nm technology of SilTerra, Inc.) and successfully tested. Figure 4f shows the layout of the new array. Its area per cell is 2.3 times larger than the original one (Fig. 4c) due to the additional real estate needed to accommodate two gate lines for each cell column.

To verify that the new array architecture enables a full inhibition of half-select disturb effects, we have performed a series of experiments, tuning all 8 cells in a 2×2 supercell array, one by one, to pre-selected goal values with a ~1% precision (Fig. 5), using a simple, fully automated feedback procedure that had been originally developed for tuning memristive devices [22, 23]. Its algorithm consists of alternating "tune" (either program or erase) and "read" pulses applied to the selected device. Every read measurement determines the necessary direction of the next tune operation, i.e. whether program or erase pulses are needed. If a read measurement shows that the desired value has been overshot, the tuning pulse polarity is changed. The tuning procedure stops when the device has reached the desired analog state with the pre-specified precision [22, 23].

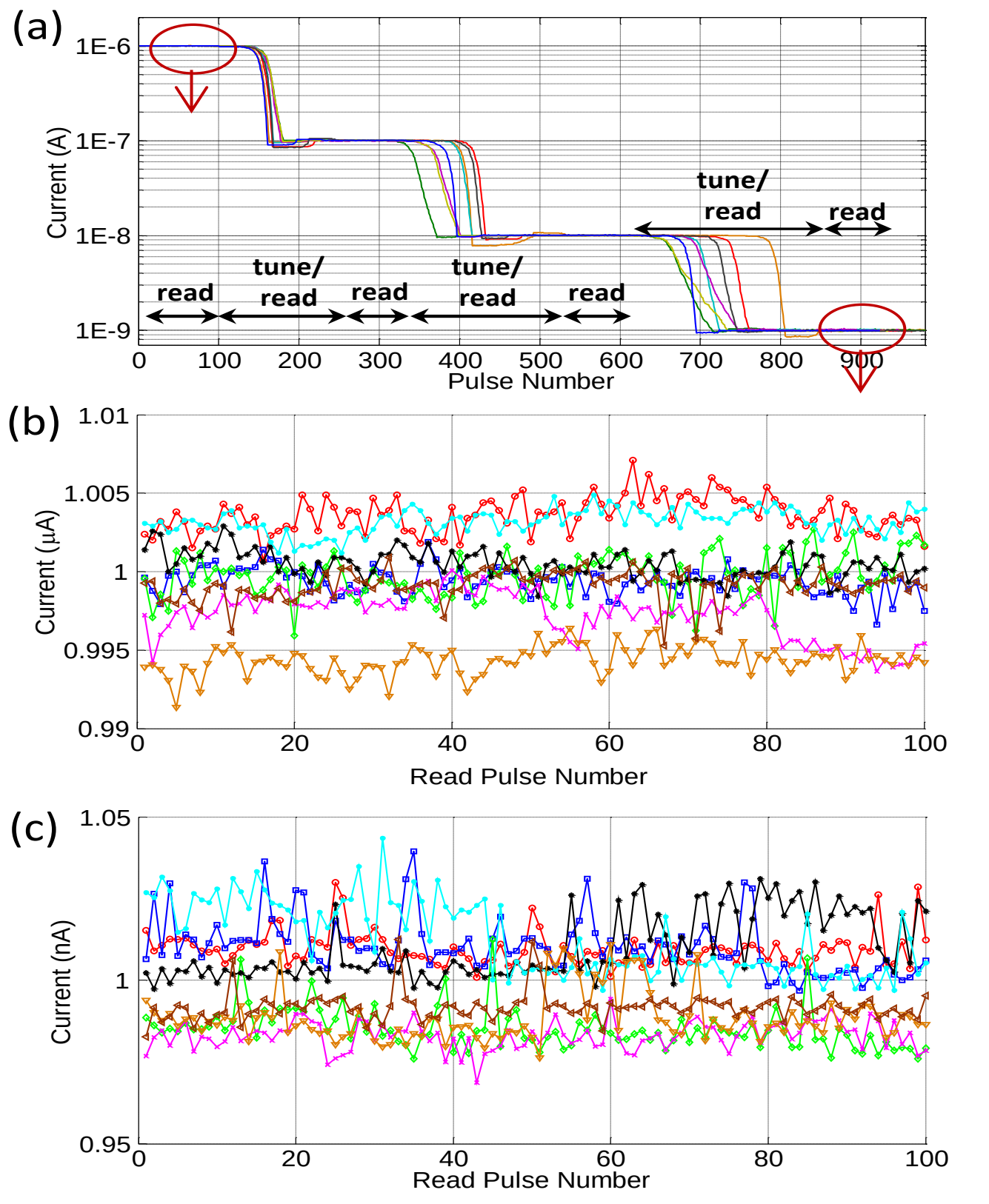

Fig. 5. High-precision tuning of cells of the modified memory: (a) All cells being tuned sequentially to 1 μA, 100 nA, 10 nA, and 1nA readout currents (as measured at $V_G$ = 2.5V, $V_D$ = 1V, $V_S$ = 0V); (b, c) zoom-in on the readout of the first and the last of the tuned states, to highlight the current variations due to intrinsic device noise. On all panels, each point represents the current average over a 10-ms time interval.

In the particular series of experiments shown in Fig. 5, the initial erase was performed with a 10-ms, 10-V gate pulse, keeping $V_D = V_S = 0$, while the initial programming, with a 5-μs, 9-V source pulse, keeping $V_D$ = 0V and $V_G$ = 1.6 V. The gradual programming was done using 5-μs source voltage pulses with an initial amplitude of 4.5 V, which was then ramped up to 8V in 50-mV steps, while applying dc voltages $V_G^P$ = 1.6V, $V_G^{P'}$ = -1V and keeping other lines grounded. The gradual erase was performed using 0.6-ms gate pulses with an initial amplitude of 5V, which was then increased to the maximum value of 8.5V, also in 50-mV steps, while applying dc voltages $V_D^E$ = 2.7V, $V_S^E$ = 0V, $V_S^{E'}$ = 2.7V, and keeping other lines grounded. (This choice of voltages is likely suboptimal and may be improved to increase tuning speed.)

We have used the high-precision tuning in the modified array for a preliminary demonstration of a small-scale four-quadrant gate-coupled vector-by-matrix multiplication [4], in which peripheral floating-gate transistors had been implemented with the same SST memory technology and integrated on the same chip. The results (Fig. 6) show an excellent linearity (derivate variation below 1%) of circuit's transfer characteristics over a wide range of input currents.

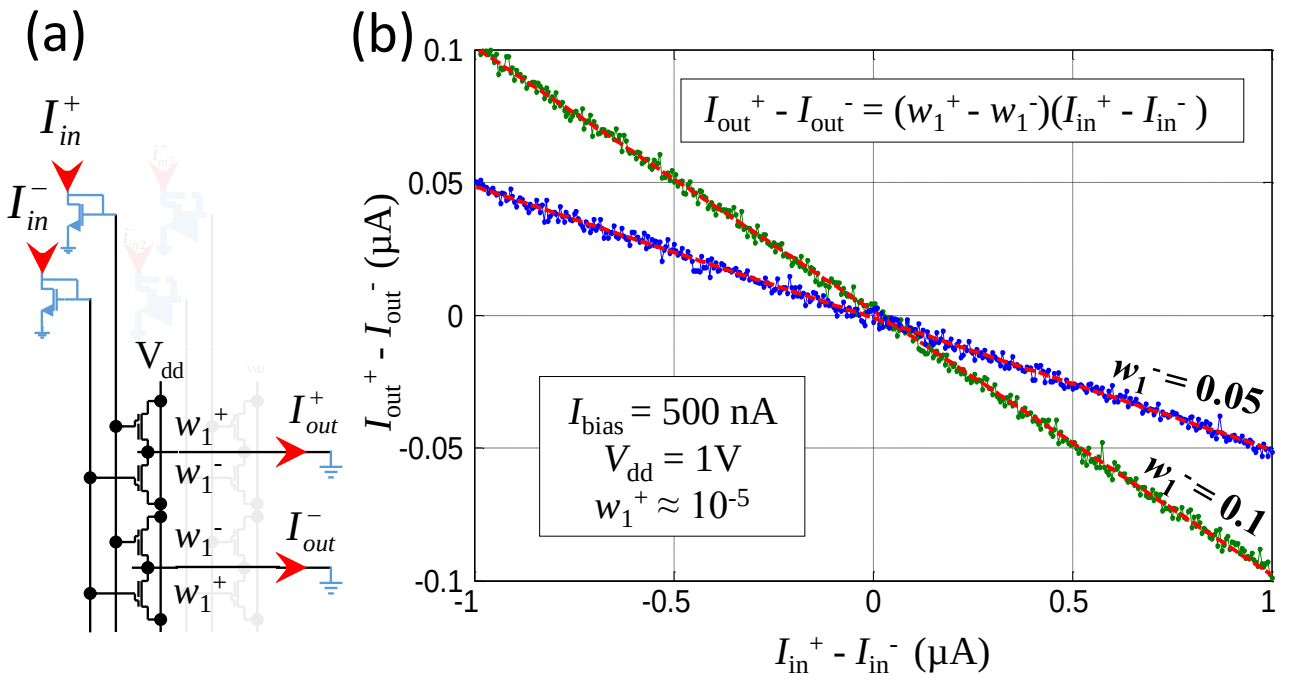

Fig. 6. Preliminary experimental results for a gate-coupled vector-by-matrix multiplier: (a) – circuit schematics and (b) measured transfer characteristics for two sets of "weights" (matrix elements) $w_1^-$. Dotted lines show another column of the array, disengaged in these experiments.

## IV. Summary and Discussion

Our initial experimental results demonstrate the possibility of high-precision, gradual, individual tuning of every floating-gate memory cell, with negligible disturb of other cells, in the modified array shown in Fig. 4d,e. For example, as Fig. 5b shows, we have been able to set the average currents of each cell within a 1% range of the target value 1 μA. As Fig. 5c shows, the tuning precision is somewhat cruder (~ 4%) for the lowest chosen target current (1 nA), but this could be expected, given the relatively larger intrinsic device noise. Though the modified array is more sparse than the original one, it is still much more compact (in terms of $F^2$) than the synaptic transistors described in any publications we are aware of. For example, it is at least 30× denser then that described in Ref. 17 (taking into account the necessary auxiliary circuitry – see p. 20 of Ref. 24), and ~500× denser in comparison with other similar cell arrays [15].

While our initial results are highly encouraging, more work is needed to prove that the floating gate device variability may allow to sustain the demonstrated high tuning precision in larger-scale cell arrays necessary for solving practical tasks of analog computation.